\documentclass[prb,preprint]{revtex4}

\pdfoutput=1

\usepackage{graphicx}
\usepackage{latexsym}
\usepackage{MnSymbol}

\begin{document}

\title{Why are stripes in $\mathbf{La_{1.88}Sr_{0.12}CuO_4}$ rotated by $\mathbf{3 ^{\circ}}$?}

\medskip 

\date{May 9, 2022} \bigskip

\author{Manfred Bucher \\}
\affiliation{\text{\textnormal{Physics Department, California State University,}} \textnormal{Fresno,}
\textnormal{Fresno, California 93740-8031} \\}

\begin{abstract}
The rotated stripes are a consequence of the orthorhombic crystal lattice and the isotropy of Coulomb repulsion between pairs of doped holes, residing at oxygen lattice sites. With stripe slanting, the  doped-hole pairs come closer to equidistance than without. The slant ratio depends on the orthorhombicity $o$ as $s = \sqrt{o/2}$.

\end{abstract}

\maketitle
\section{ROTATED STRIPES}
Rotated  stripes of both magnetic and charge order, slightly ($\sim3^\circ$) slanted from the direction of the copper-oxygen bonds in the $CuO_2$ planes of $La_{1.88}Sr_{0.12}CuO_4$, have been observed, some time ago, with neutron scattering\cite{1} and X-ray diffraction,\cite{2} respectively. In the low-temperature phases of doped lanthanum cuprates, $La_{2-x}Ae_xCuO_4$ ($Ae = Sr, Ba$), the $CuO_6$ octahedra are tilted, which gives rise to monoclinic lattice distortions.\cite{3,4,5,6,7,8,9,10} As a result, the magnetic and charge order from $Ae$-doping appear as unidirectional ``stripes'' rather than a checkerboard pattern, with domains of $a$-stripes and $b$-stripes. The deviation from tetragonality 
is characterized by the relative difference of the planar lattice constants,
\begin{equation}
    o \equiv 2\; \frac{b_0 - a_0 }{b_0 + a_0 }\;,
\end{equation}
called ``orthorhombic strain'' or ``orthorhombicity,'' for short.
It has been suspected that the rotated stripes of $La_{1.88}Sr_{0.12}CuO_4$ in the low-temperature orthorhombic (LTO) phase arise due to the crystal's orthorhombicity.\cite{1,2} On the atomic scale, the slanted stripes can be regarded as kinked rows and columns of atomic entities parallel and perpendicular to the directions of the copper-oxygen bonds in the host crystal.\cite{1,2,11,12} Recently, an explanation for the rotated stripes has been given with a doped Hubbard model in terms of an anisotropic electron hopping amplitude, $t'$, between next nearest neighbors.\cite{11} The present paper presents a simpler explanation, based on (i) the orthorhombicity of the host crystal, (ii) the assumption that the magnetic and charge order both arise from a superlattice of $O$ atoms, hosting pairs of doped holes and residing at oxygen lattice sites,\cite{13} and (iii) the isotropy of Coulomb repulsion between them (being positively charged relative to the host crystal), promoting equidistance.

\includegraphics[width=3.25in]{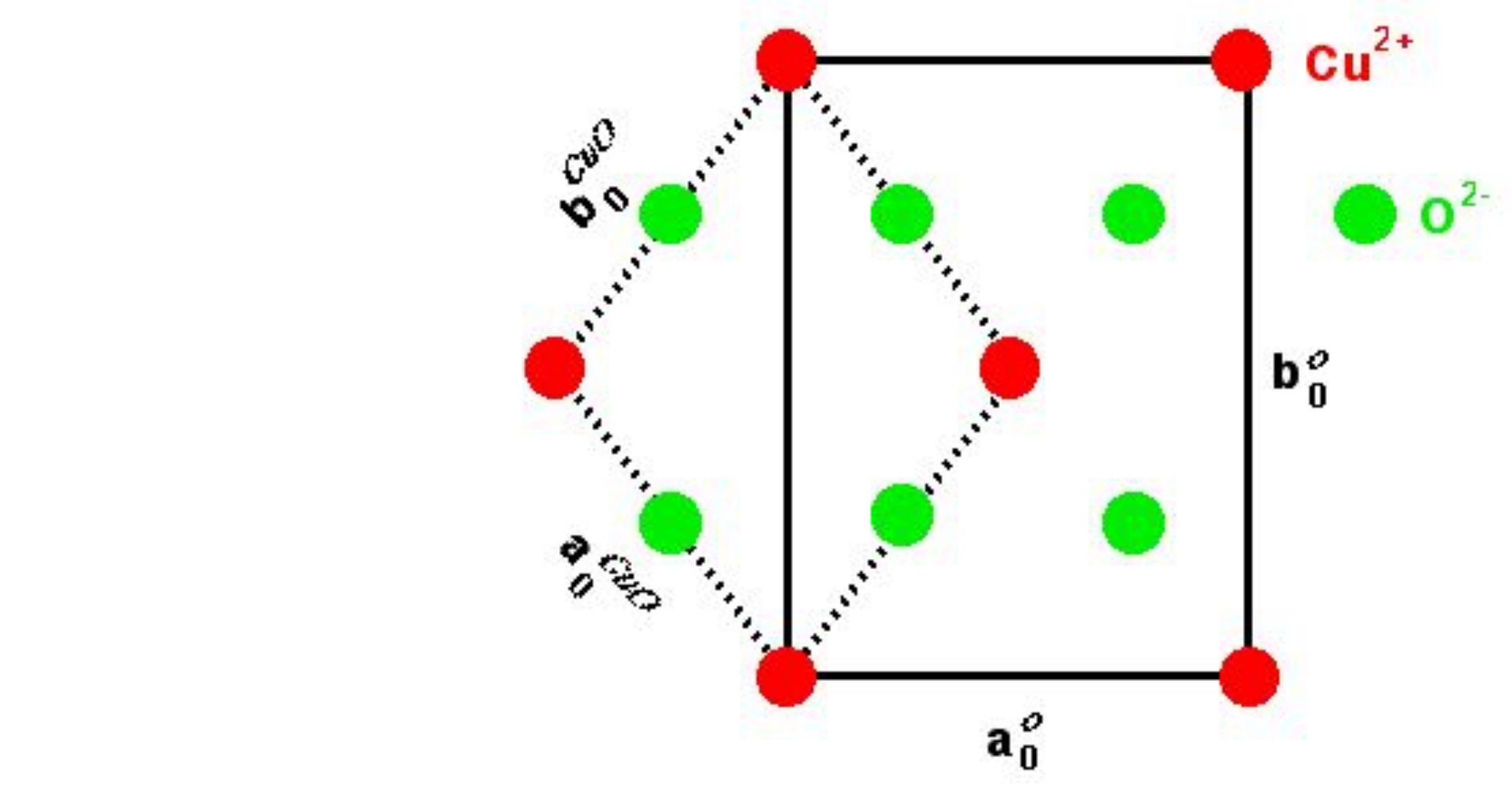}

\footnotesize \noindent FIG. 1. Orthorhombic unit cell $a_0^o \times b_0^o$ of the LTO phase of $La_{1.88}Sr_{0.12}CuO_4$ (solid line), and unit cell $a_0^{CuO} \times b_0^{CuO}$ along the copper-oxygen bonds, called ``$CuO_2$ plaquette'' (hatched line).  The orthorhombicity is much exaggerated for clarity (here, $o = 30 \%$). \normalsize

\section{LATTTICES FOR THE LTO PHASE}  
The LTO phase of $La_{1.88}Sr_{0.12}CuO_4$ is orthorhombic, with planar lattice constants $a_0^o = 5.3184$ \AA $\;$ and $b_0^o = 5.3450$ \AA, amounting, by Eq. (1), to an orthorhombicity $o = 0.5 \%$.\cite{1} For a comparison of scattering or diffraction peaks, it is convenient to assign planar lattice axes along the copper-oxygen bonds, $a^{CuO}$ and $b^{CuO}$, as in the high-temperature tetragonal (HTT) phase. The convenience comes at a price: The LTO  $a_0^{CuO} \times b_0^{CuO}$ unit cell is an equilateral \textit{skewed} rhomboid rather than a rectangle (see Fig. 1). Two options of \textit{approximate} Cartesian coordinate systems (that is, with \textit{orthogonal} axes), rotated from the orthorhombic axes ($a^o, b^o$) by $45 ^\circ$, are either using quasi-tetragonal notation,\cite{14} with axes $a^{qt}$ and $b^{qt}$ and lattice constants $a_0^{qt}=b_0^{qt} = (a_0^o + b_0^o)/(2 \sqrt{2})$, or using quasi-orthogonal notation with axes $a^{qo}$ and $b^{qo}$ and lattice constants $a_0^{qo} = a_0^o/\sqrt{2}$ but $b_0^{qo} = b_0^o/\sqrt{2} = (1+o)a_0^{qo}$. 

Each option has advantages and disadvantages---neither one is exact. Their choice depends on the purpose. Quasi-tetragonal notation is preferable for comparisons of scattering or diffraction experiments: Although the $a^{qt}$ and $b^{qt}$ axes don't account for orthorhombicity, effects from the latter appear as rotated peaks in reciprocal space.\cite{1} On the other hand, quasi-orthorhombic notation is better suited for expressing stripe slanting in the crystal, particularly under uniaxial stress. The first notation, with its tetragonal axes, diverts attention from the cyrstal's orthorhombicity. The second notation, with its orthorhombic axes, emphasizes orthorhombicity.

\section{DEPENDENCE OF THE SLANT ON ORTHORHOMBICITY}
A planar distribution of equal point charges, located on a quasi-orthorhombic lattice with spacing
\begin{equation}
    B_0 = (1+o)A_0 \;,
\end{equation}
as in Fig. 2, is \textit{unstable}, due to its deviation from equidistance ($B_0 > A_0$), and will rearrange under slight disturbance. To this end we consider the rectangular (blue) frame $S \times T$  with 
\noindent side $S$ and top $T$,
\begin{equation}
    S= nB_0\;,\;\;\;\;T= nA_0 \;.
\end{equation}
The number $n$ is determined by the condition that the (red) straight line $T''$, from the top left corner of the rectangle $S \times T$, is rotated by a slant

\includegraphics[width=6in]{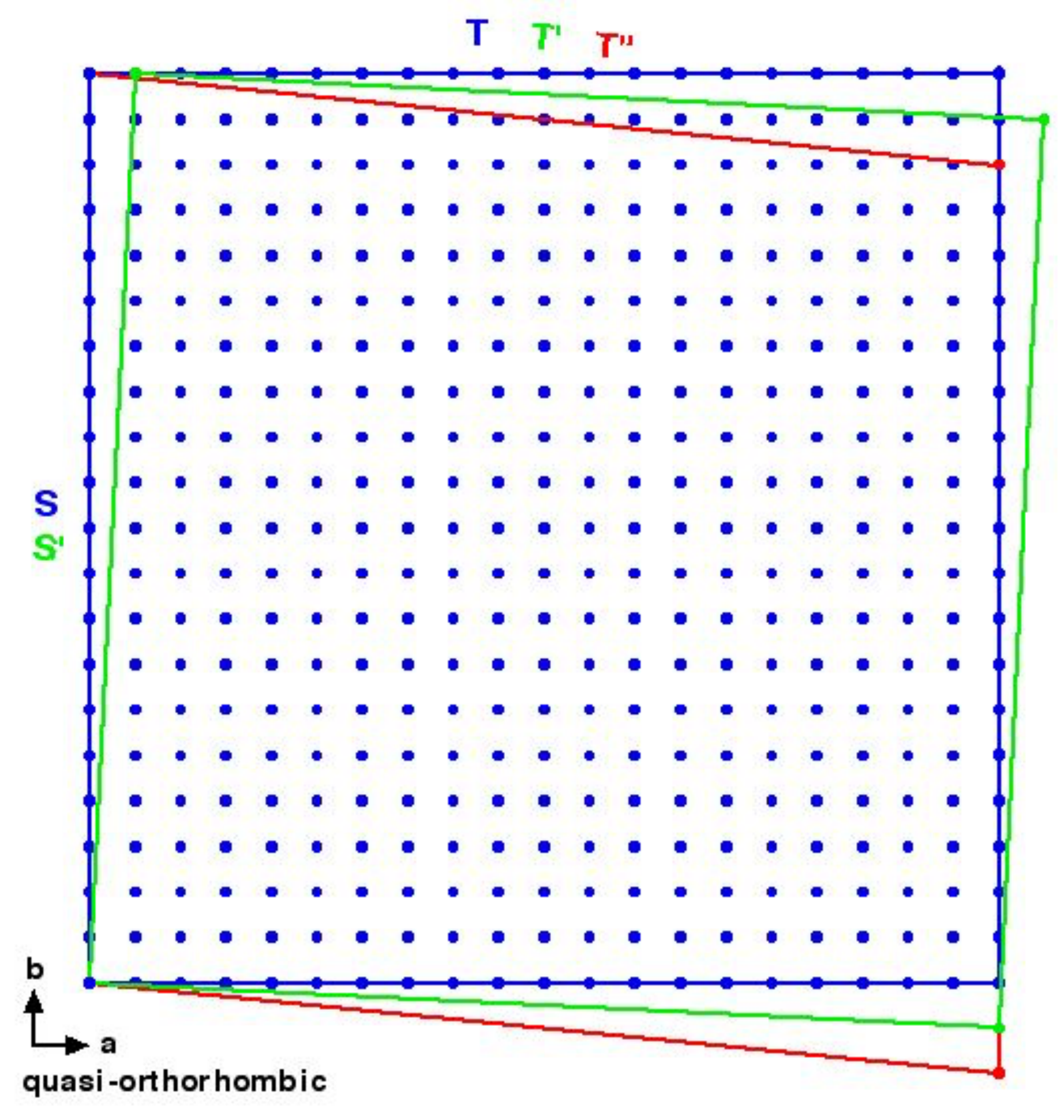}  \footnotesize

\noindent FIG. 2. Quasi-orthorhombic lattice of equal point charges (blue dots) with lattice constants $A_0$ and $B_0 = (1+o)A_0$: Only an $n \times n$ rectangle of the infinite $a \times b$ plane is shown, bounded by the (blue) frame $S \times T$ with top $T$ and side $S = (1+o)T$ (here, $o = 0.5 \%, \; n = 20$). Lacking equidistance between the point charges, the lattice is in an \textit{unstable} equilibrium and will rearrange upon slight disturbance. The number $n$ is determined by the condition that the (red) top $T''$ is rotated such that its far end is laterally \textit{two} spacings away from that of $T$ \textit{and} is of length $T'' = S$. This makes the (blue-red) rhomboid $S \times T''$ equilateral---but still unsuitable for equidistance because of different diagonal lengths. Treating all boundaries of the frame on an equal footing, equidistance can be achieved by confining the point charges into the (green) square $S' \times T'$ with $S' = T' \simeq (S+T)/2$, rotated by a slant of  $s = 1/n$. \normalsize \pagebreak 

\begin{equation}
    s_{T''} = \frac{2}{n} \;,
\end{equation}
until its far end is \textit{two} superlattice spacings laterally away from that of $T$ \textit{and} that its length equals the side of the rectangle,
\begin{equation}
  \sqrt{n^2 +2^2 (1+o)^2} A_0 =  T'' \;\; = \;\;\; S = (1+o)nA_0 \;.
\end{equation}
This is illustrated in Fig. 2 for the case with $o = 0.5 \%$ and $n=20$.
Although equilateral, the resulting (blue-red) frame $S \times T''$ is unsuitable for equidistant spacing of included point charges because of unequal diagonal lengths. The point charges at the corners of its rhomboid unit cells, $A_0 \times B''_0$, have equal nearest-neighbor (NN) distances, but unequal next-nearest-neighbor (NNN) distances diagonally, which makes the lattice unstable. Avoiding preference of one lattice axis over the other, we instead transform the rectangle $S \times T$
to the (green) square $S' \times T'$  with edge lengths $S' = T' \simeq (T + T'')/2 = (S + T)/2$ and slant
\begin{equation}
    s \equiv s_{S'} = s_{T'} = \frac{1}{2}s_{T''} =  \frac{1}{n} \;.
\end{equation}
From Eq. (5) we have
\begin{equation}
    n^2 + 4(1 +2o +o^2) =n^2 + 4 +8o +4o^2 \;\;\; = \;\;\; (1+o)^2n^2 = n^2+2on^2+o^2n^2 \;.
\end{equation}
After rearrangement and neglect of the relatively small $o^2$ terms we maintain 
\begin{equation}
    2(1+2o)  =  on^2 \;.
\end{equation}
With further neglect of $2o \ll 1$, we obtain the dependence of the slant on the orthorhombicity,
\begin{equation}
  s =  \frac{1}{n}  =  \sqrt{\frac{o}{2}}  \;.
\end{equation}
In the present case, with $o = 0.005 = 1/200$, this gives $s = 0.05 = 1/20$.

\section{SLANTED SUPERLATTICE}
Figure 3 shows a rotated charge order—a slanted superlattice of $O$ atoms—in the $CuO_2$ plane of (here, very slight) orthorhombic symmetry ($o = 0.005$), rotated by a slant $s=1/20$ from the quasi-orthorhombic axes $(a^{qo}, b^{qo})$ that are approximately along the copper-oxygen bonds. Generally, the spacing of the superlattice is given by the reciprocal of the incommen- 

\includegraphics[width=5.95in]{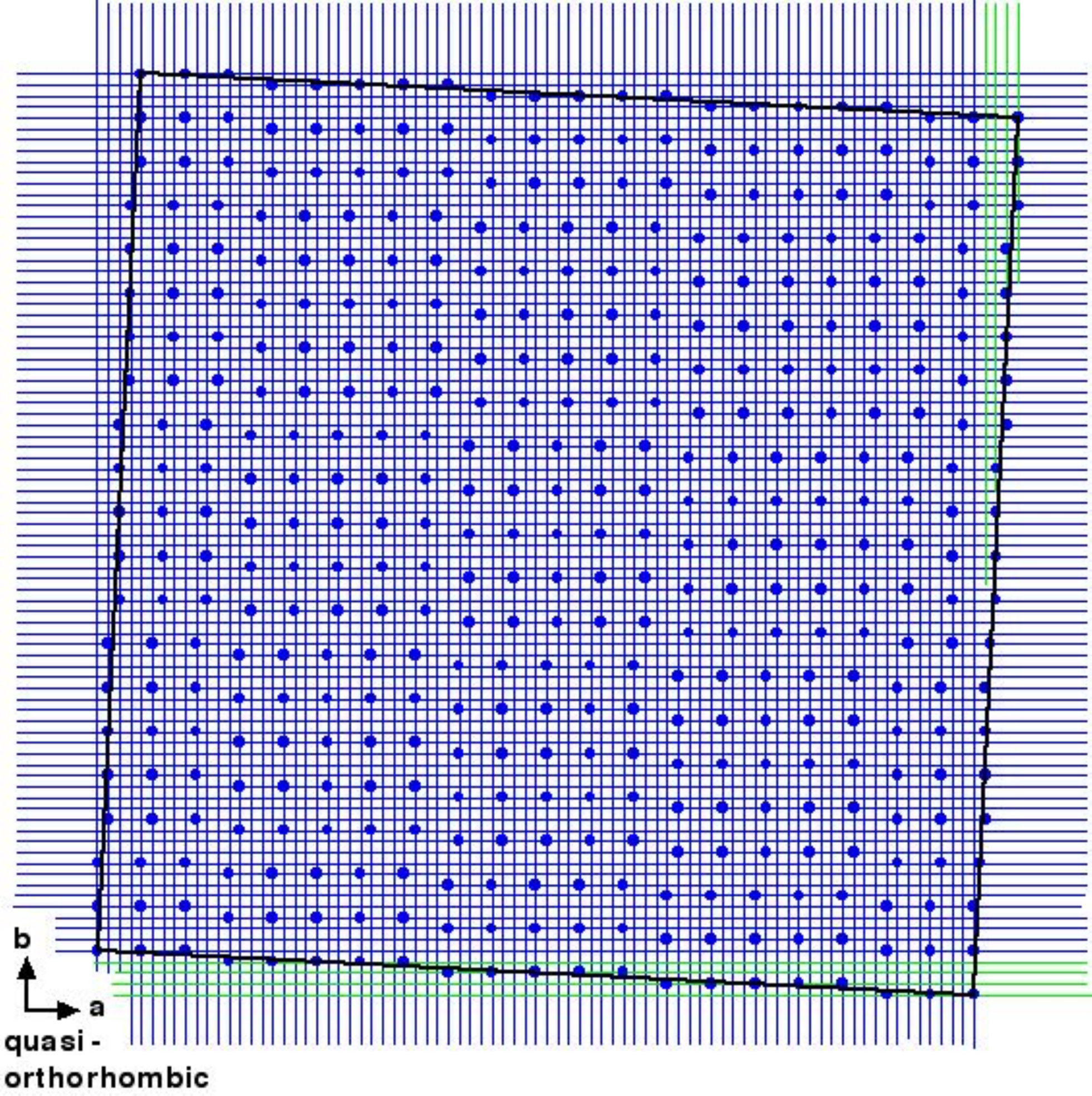} \footnotesize

\noindent FIG. 3. Realistic rendition of the period-four superlattice of $O$ atoms (blue dots) in the $CuO_2$ plane of $La_{2-x}Sr_xCuO_4$ ($x \approx 1/8$) in the LTO phase, rotated from the quasi-orthorhombic axes ($a^{qo}, b^{qo}$) by a slant $s=1/20$ due to the orthorhombicity $o = 0.5 \%$. Oxygen lattice sites of the crystal are at the intersections of the fine grid lines, copper lattice sites are between. The superlattice spacing is $A_0^{qo} = 4a_0^{qo}$ and $B_0^{qo} = 4b_0^{qo}$. In the rotated superlattice the pairs of doped holes must reside in $O$ atoms at oxygen sites, but they can choose at \textit{which} one, such that they come closest to equidistance. The result are jagged columns of $O$ atoms—here $16 = 4 \times 4$ crystal lattice spacings along a quasi-orthorhombic axis, followed by a forward skip of four lattice spacings and a sideways jog of one. Together, their 16-unit length and 4-unit skip amount to a kink period of 20 lattice spacings. The frame's
macroscopic slant, $\sigma = 1/20$, is the average over the kinked atomic $O$ columns. The rotation achieves a compromise of three requirements: (i) orthorhombic lattice, (ii) $O$ \linebreak
atoms residing only at oxygen sites, and (iii) equidistant distribution of $O$ atoms due to Coulomb repulsion.  \normalsize

\noindent surability of charge-order stripes, $L_0^{co} = 1/q^{co}_{\parallel}$. Its exact value in $La_{1.88}Sr_{0.12}CuO_4$ is $L_0^{co} =  1/(0.236$ r.l.u.)= $4.24 \; a_0$, being somewhat larger than four lattice spacings.\cite{1,2} For the sake of simplicity, the closely related \textit{commensurate}, period-four superlattice of $L_0^{co} = 4 a_0$ is shown in Fig. 3, resulting from $Sr$-doping $x\simeq1/8$. Isotropic Coulomb repulsion between $O$ atoms (of positive charge relative to the host crystal) makes them occupy every fourth oxygen lattice site. Each tiny square in the figure corresponds to a $CuO_2$ plaquette. Oxygen lattice sites are at the intersection of the fine grid. Host lattice ions, $Cu^{2+}$ and $O^{2-}$, are not shown for clarity.
The rotation of the superlattice achieves a compromise between the three stated conditions: orthorhombicity, lattice-site residence, and equidistance (resulting from mutual Coulomb repulsion). While the doped-hole pairs must reside at oxygen sites of the orthorhombic crystal lattice, they can choose at \textit{which} of those sites, such that they come closest to equidistance with their $O$ neighbors (strictly speaking, to zero net Coulomb force from all their $O$ neighbors). 

It seems worth pointing out that the same assumptions---equidistance of pairs of doped holes or electrons (at $O$ or $Cu$ atoms) due to Coulomb repulsion---were used (in this case based on a tetragonal lattice) in the derivation of the doping dependence of magnetic ($m$) and charge-order ($c$) stripe incommensurability, $q^{\kappa}_{\parallel}(x)$, $\kappa = m,c$, of hole-doped and electron-doped `$214$' cuprates.\cite{13}

\section{STRIPE ROTATION UNDER UNIAXIAL STRESS}
Very recently\cite{12} uniaxial pressure was applied to $La_{1.88}Sr_{0.12}CuO_4$ along one of the directions of the $CuO$ bonds.
Using quasi-tetragonal notation, the direction was assigned $[010]_{qt}$  (i. e. along $b^{qt}$), amounting to in-plane compressive strain, $\epsilon_b = (b - b_0)/b_0$, with three strain conditions: $\epsilon_{b,0} = 0 < |\epsilon_{b,1}| < |\epsilon_{b,2}| < 0.04 \%$. It caused a lesser transverse incommensurability $q^c_{\perp}$ at strain $\epsilon_{b,1}$ and its disappearance, $q^c_{\perp}=0$, at strain $\epsilon_{b,2}$, with corresponding decrease of  slant ratio $s$ and rotation angle $\theta$, accompanied by an increase of the kink length $n$ (see Table I). Qualitatively, the reduced rotation angle $\theta$ under strain $\epsilon_b$ can be regarded as a consequence of reduced orthorhombicity of the crystal. For a quantitative treatment the question arises as to which approximate notation to use: quasi-tetragonal or quasi-orthorhombic? The choice can be decided by future experiments where the crystal is instead subjected to uniaxial compressive strain $\epsilon_a$ in the \textit{other} direction of $CuO$ bonds. Quasi-tetragonal notation, with equivalent $(a^{qt}, b^{qt})$ axes, is appropriate if $\epsilon_a$ likewise causes a reduction of  stripe rotation. However, if $\epsilon_a$ causes an \textit{increase} of the rotation angle $\theta$, reflecting increased orthorhombicity, then quasi-orthorhombic notation is called for.

\begin{table}[ht]

\begin{tabular}{|p{2.6cm}|p{2.2cm}|p{1.8cm}|p{1.2cm}|p{1cm}|p{1.8cm}|} \hline  \hline
 
$\;\;$Strain &$\;\;q^c_{\perp}\;\;(r.l.u.)$&$\;\;\;\;\;\;s$&$\;\;\;\;n$&$\;\;\;\;\theta$&$\;\;\;\;\;\;o $\\
 \hline  \hline

$\;\;\epsilon_{b,0} = 0$&$\;\;\;\;0.011 $&$\;\;\;0.051 $&$\;\;\;20 $&$\;\;\;3^\circ$&$\;\;\;0.005 $\\ \hline

$\;\;\epsilon_{b,1}  $&$\;\;\;\;0.007 $&$\;\;\;0.031 $&$\;\;\;32 $&$\;\;\;2^\circ$&$\;\;\;0.002^* $\\ \hline 
$\;\;|\epsilon_{b,2}|<0.04\%$&$\;\;\;\;0 $&$\;\;\;0 $&$\;\;\; \infty $&$\;\;\; 0 $&$\;\;\; 0^* $\\ \hline
 \hline
\end{tabular}
\footnotesize \caption{Transverse incommensurability $q^c_{\perp}$ of  charge-order stripes in $La_{1.88}Sr_{0.12}CuO_4$, slant ratio $s = q^c_{\perp}/q^c_{\parallel}=1/n$  of kinked stripes, kink length $n$ (in crystal lattice spacing), rotation angle $\theta = \arctan s$, 
and orthorhombicity $o$ without and with compressive strain (Ref. 12). The longitudinal incommensurability is $q^c_{\parallel} = 0.236$ r.l.u. (Ref. 2). \;\;\;\;\;\;\;$^*$calculated with Eq. (9)} \normalsize
\label{table:1}
\end{table}

\section{DOMAIN POPULATION UNDER UNIAXIAL STRESS}
Uniaxial pressure on $La_{1.88}Sr_{0.12}CuO_4$ causes not only stripe rotation but, as recently discovered,\cite{15,16} also affects the tilt of $CuO_6$ octahedra that gives rise to domains of $a$-stripes or $b$-stripes, as mentioned above. Those experiments shed more light on the question of stripe rotation under $a$- \textit{vs.} $b$-strain.
Without uniaxial stress, domains of $a$-stripes and of $b$-stripes, are equally populated. For charge-order stripes this is shown by X-ray Bragg peaks of equal height in $(h,0,12.5)$ and $(0,k,12.5)$ scans.\cite{15} However, under uniaxial $b$-axis pressure, $|\epsilon_b| < 0.04 \%$, the $k$-peak disappears, whereas the $h$-peak doubles in height. This is interpreted as a depopulation of the $b$-stripe domains in favor of $a$-stripe domains.

In a similar experiment with neutron scattering off magnetic stripes in $La_{1.88}Sr_{0.12}CuO_4$, the height of the $(h,\frac{1}{2},0)$ Bragg peak and the equivalent $(-\frac{1}{2},k,0)$ peak were found equal when no uniaxial pressure was applied.\cite{16} Under $a$-axis strain, $|\epsilon_a| \lessapprox -0.02 \%$, the $h$-peak disappears, whereas the $k$-peak doubles its height. This shows a depopulation of the $a$-stripe domains in favor of $b$-stripe domains. A common result of the X-ray and neutron scattering experiments, uniaxial pressure along one planar axis depopulates charge-order and magnetic stripes of the eponymous domains in favor the perpendicular ones. 

One then would expect that uniaxial \textit{tension} has the opposite effect, namely an increased signal from stripes along the direction of the applied tension. This is indeed observed for $La_{1.475}Nd_{0.4}Sr_{0.125}CuO_4$.\cite{9} When unbiased, the crystal is in the LTT phase. Application of tension along the $a$-direction breaks  tetragonality, as indicated by a lower phase-transition temperature $T_{LTT}$, and an increase of the signal of $a$-oriented charge stripes. 

A difference between the compressive uniaxial $b$-strain and $a$-strain experiments\cite{12,16} concerns the transverse incommensurability $q^{\kappa}_{\perp}$ (also called $Y$-shift\cite{1}),  with $\kappa = c, m$ for charge-density and magnetic stripes, respectively. As Table II shows, stripe rotation disappears under $b$-strain (treated in Sect. V), but \textit{persists} under $a$-strain. In the latter case, stripe 
rotation does not increase, as quasi-orthorhombic notation would predict, but it certainly does not disappear. More experiments on charge order and magnetization order in $La_{1.88}Sr_{0.12}CuO_4$ under uniaxial (compressive and tensile) stress should bring more clarification.

\begin{table}[ht]

\begin{tabular}{|p{2.6cm}|p{0.8 cm}| p{2.2cm}|p{1.8cm}|p{1.2cm}|p{1cm}|p{1.8cm}|} \hline  \hline
 
$\;\;$Strain &$\;\;\;\kappa$&$\;\;\;q^{\kappa}_{\perp}\;\;$(r.l.u.)&$\;\;\;\;\;\;s$&$\;\;\;\;n$&$\;\;\;\;\theta$&$\;\;\;\;\;\;o $\\
 \hline  \hline

$\;\;\epsilon_{b,0} = 0$&$\;\;\;$c&$\;\;\;\;0.011 $&$\;\;\;0.051 $&$\;\;\;20 $&$\;\;\;3^\circ$&$\;\;\;0.005 $\\ \hline

$\;\;|\epsilon_b|<0.04\%$&$\;\;\;$c&$\;\;\;\;0 $&$\;\;\;0 $&$\;\;\; \infty $&$\;\;\; 0 $&$\;\;\; 0^* $\\ \hline \hline

$\;\;\epsilon_{a,0} = 0$&$\;\;\;$m&$\;\;\;\;0.007 $&$\;\;\;0.059 $&$\;\;\;17 $&$\;\;\;3^\circ$&$\;\;\;0.005 $\\ \hline

$\;\;|\epsilon_{a}| <\approx 0.02\%$&$\;\;\;$m&$\;\;\;\;0.005 $&$\;\;\;0.043 $&$\;\;\;24 $&$\;\;\;2^\circ$&$\;\;\;0.0036^* $\\ \hline \hline 

\end{tabular}
\footnotesize \caption{Transverse incommensurability $q^{\kappa}_{\perp}$ of charge-order stripes ($\kappa = c$) and magnetic stripes ($\kappa = m$) in $La_{1.88}Sr_{0.12}CuO_4$, slant ratio $s = q^{\kappa}_{\perp}/q^{\kappa}_{\parallel} = 1/n$ of kinked stripes, kink length $n$ (in crystal lattice spacing), rotation angle $\theta$, and orthorhombicity $o$ without and with compressive strain (Refs. 12, 16). The longitudinal incommensurabilities are
$q^c_{\parallel} = 0.236$ r.l.u. and $q^m_{\parallel} = 0.118$ r.l.u. (Refs. 2, 1).
\;\;\;\;\;\;\;\;\;\;\;\;\;\;\;\;\;\;\;\;\;\;\;\;\;\;\;\;\;\;\;\;\;\;\;\;\;\;\;\;\;\;\;\;\;\;\;\;\;\;\;\;\;\;\;\;\;\;\;\;\;\;\;\;\;\;\;\;\;\;\;\;\;\;\;\;\;\;\;\;\;\;\;\;\;\;\;\;\;\;\;\;\;\;\;\;$^*$calculated with Eq. (9)} \normalsize

\label{table:2}
\end{table}

\centerline{ \textbf{ACKNOWLEDGMENT}}

\noindent I thank Qisi Wang for clarifying correspondence.


\begin{thebibliography}{16}

\bibitem{1} H. Kimura, H. Matsushita, K. Hirota, Y. Endoh, K. Yamada, G. Shirane, Y. S. Lee, M. A. Kastner, and R. J. Birgeneau: Incommensurate geometry of the elastic magnetic peaks in superconducting $La_{1.88}Sr_{0.12}CuO_4$, Phys. Rev. B \textbf{61}, 14366 (2000).

\bibitem{2} V. Thampy, M. P. M. Dean, N. B. Christensen, L. Steinke, Z. Islam, M. Oda, M. Ido, N. Momono, S. B. Wilkins, and J. P. Hill: 
Rotated stripe order and its competition with superconductivity in $La_{1.88}Sr_{0.12}CuO_4$, Phys. Rev. B \textbf{90}, 100510(R) (2014).

\bibitem{3} M. Reehuis, C. Ulrich, K. Proke\v{s}, A. Gozar, G. Blumberg, S. Komiya, Y. Ando, P. Pattison, and B. Keimer: Crystal structure and high-field magnetism of $La_2CuO_4$, Phys. Rev. B \textbf{73}, 144513 (2006).

\bibitem{4} J. A. Robertson, S. A. Kivelson, E. Fradkin, A. C. Fang, and A. Kapitulnik: Distinguishing patterns of charge order: Stripes or checkerboards, Phys. Rev. B textbf{74}, 134507 (2006).

\bibitem{5} M. H\"{u}cker, M. v. Zimmermann, G. D. Gu, Z. J. Xu, J. S. Wen, G. Xu, H. J. Kang, A. Zheludev, and J. M. Tranquada: Stripe order in superconducting $La_{2-x}Ba_xCuO_4$ ($0.095 \le x \le 0.155$). Phys. Rev. B \textbf{83}, 104506 (2011).

\bibitem{6} A. J. Achkar, M. Zwiebler, Ch. McMahon, F. He, R. Sutarto, I. Djianto, Z. Hao, M. J. P. Gingras, M. H\"{u}cker, G. D. Gu, A. Revcolevschi, H. Zhang, Y.-J. Kim, J. Geck, and D. G. Hawthorn:
Nematicity in stripe-ordered cuprates probed via resonant x-ray scattering, Science \textbf{351}, 576 (2016).

\bibitem{7} Anar Singh, J. Schefer, R. Sura, K. Conder, R. F. Sibille, M. Ceretti, M. Frontzek, and W. Paulus: Evidence for monoclinic distortion in the ground state phase of underdoped
$La_{1.95}Sr_{0.05}CuO_4$: A single crystal neutron diffraction study, J. Appl. Phys. \textbf{119}, 123902 (2016).

\bibitem{8} A. Sapkota , T. C. Sterling , P. M. Lozano, Y. Li , H. Cao , V. O. Garlea, D. Reznik Q. Li, I. A. Zaliznyak, G. D. Gu, and J. M. Tranquada: Reinvestigation of crystal symmetry and fluctuations in $La_2CuO_4$, Phys. Rev. B \textbf{104}, 014304 (2021).

\bibitem{9} T. J. Boyle , M. Walker, A. Ruiz, E. Schierle, Z. Zhao, F. Boschini, R. Sutarto, T. D. Boyko, W. Moore, N. Tamura, F. He, E. Weschke, A. Gozar, W. Peng, A. C. Komarek, A. Damascelli, 
C. Sch\"{u}\ss ler-Langeheine, A. Frano, E. H. da Silva Neto, and S. Blanco-Canosa: Large response of charge stripes to uniaxial stress in $La_{1.475}Nd_{0.4}Sr_{0.125}CuO_4$, Phys. Rev. Res. \textbf{3}, L022004 (2021).

\bibitem{10} R. Frison, J. K\"{u}spert, Q. Wang O. Ivashko, M. v. Zimmermann, M. Meven, D. Bucher, J. Larsen, 6 Ch. Niedermayer, M. Janoschek, T. Kurosawa, N. Momono, M. Oda, N. B. Christensen, and J. Chang: Crystal Symmetry of Stripe Ordered $La_{1.88}Sr_{0.12}CuO_4$, arXiv:2201.04858

\bibitem{11} W. He, J. Wen, H.-C. Jiang, G. Xu, W. Tian, T. Taniguchi, Y. Ikeda, M. Fujita, and Y. S. Lee: Prevalence of tilted stripes in $La_{1.88}Sr_{0.12}CuO_4$ and the importance of $t$' in the Hamiltonian, arXiv:2107.10264

\bibitem{12} Q. Wang, K. v Arx, D. G. Mazzone, S. Mustafi, M. Horio, J. K\"{u}spert, J. Choi, D. Bucher, H. Wo, J. Zhao, W. Zhang, T. C. Asmara, Y. Sassa, M. M\r{a}nsson, N. B. Christensen, M. Janoschek, T. Kurosawa, N. Momono, M. Oda, M. H. Fischer, T. Schmitt, and J. Chang: Uniaxial Pressure Induced Stripe Order Rotation in $La_{1.88}Sr_{0.12}Cu_O4$, arXiv:2203.09987
 
\bibitem{13} M. Bucher: Stripes in heterovalent-metal doped cuprates,    arXiv:2002.12116v9

\bibitem{14} S. Wakimoto, G. Shirane, Y. Endoh, K. Hirota, S. Ueki, K. Yamada, R. J. Birgeneau, M. A. Kastner, Y. S. Lee, P. M. Gehring, and S. H. Lee: Observation of incommensurate magnetic correlations at the lower critical concentration for superconductivity in $La_{2-x}Sr_xCuO_4$ ($x=0.05$),  Phys. Rev. B \textbf{60}, R769 (1999).

\bibitem{15} J. Choi, Q. Wang, S. J\"{o}hr, N. B. Christensen, J. K\"{u}spert, D. Bucher, D. Biscette, M. H\"{u}cker, T. Kurosawa, N. Momono, M. Oda, O. Ivashko, M. v. Zimmermann, M. Janoschek, and J. Chang: Disentangling Intertwined Quantum States in a Prototypical Cuprate Superconductor, arXiv:2009.06967

\bibitem{16} G. Simutis, J. K\"{u}spert, Q. Wang, J. Choi, D. Bucher, M. Boehm, F. Bourdarot, M. Bertelsen, C. N. Wang, T. Kurosawa, N. Momono, M. Oda, M. M\r{a}nsson, Y. Sassa M. Janoschek, N. B. Christensen, J. Chang, and D. G. Mazzone: Single-domain stripe order in a high-temperature superconductor, arXiv:2204.02304

\end{thebibliography}
\end{document}